\documentclass[9pt,twocolumn,twoside]{pnas-new}

\templatetype{pnasmathematics} 

\newcommand{\avg}[1]{\left< #1 \right>} 
\newcommand{\ket}[1]{\left| #1 \right>} 
\newcommand{\bra}[1]{\left< #1 \right|} 
\newcommand{\braket}[2]{\left< #1 \vphantom{#2} \right|
 \left. #2 \vphantom{#1} \right>} 

\title{Relativistic treatment of the energy shifts caused by static electromagnetic effects on free electrons}

\author[a,b,1]{P. Kurian}

\affil[a]{Quantum Biology Laboratory, Howard University, Washington, DC 20059}
\affil[b]{Department of Medicine, Howard University College of Medicine, Washington, DC 20059}

\leadauthor{Kurian} 

\significancestatement{Free electron systems can be found throughout nature. Weak magnetic fields acting on free electrons have demonstrated intriguing effects, including conversion of electron spin into orbital angular momentum to produce so-called electron vortex beams. Here we provide a quantum field theoretic analysis to explain why such observed phenomena are possible. We show in this article how weak, static electromagnetic fields can exert profound influences on free (and quasi-free) electron systems. In particular, starting from the Dirac equation, we demonstrate that the expectation value of the maximum energy shift induced by such fields can be orders of magnitude larger than that predicted in quantum mechanics by the Zeeman effect.}

\authorcontributions{PK completed all derivations and calculations, with numerical estimates, produced the figure, and wrote the paper.}
\authordeclaration{We have no conflict of interest.}
\correspondingauthor{\textsuperscript{1}To whom correspondence should be addressed. E-mail: pkurian@howard.edu}


\begin{abstract}
Magnetic effects on free electron systems have been studied extensively in the context of spin-to-orbital angular momentum conversion. Starting from the Dirac equation, we derive a fully relativistic expression for the energy of free electrons in the presence of a spatiotemporally constant, weak electromagnetic field. The expectation value of the maximum energy shift, which is completely independent of the electron spin-polarization coefficients, is computed perturbatively to first order. This effect is orders of magnitude larger than that predicted by the quantum mechanical Zeeman shift. We then show, in the non-relativistic limit, how to discriminate between achiral and completely polarized states and discuss possible mesoscopic and macroscopic manifestations of electron spin states across many orders of magnitude in the physical world.
\end{abstract}

\dates{This manuscript was compiled on \today}
\doi{\url{www.pnas.org/cgi/doi/10.1073/pnas.XXXXXXXXXX}}

\begin{document}

\maketitle
\thispagestyle{firststyle}
\ifthenelse{\boolean{shortarticle}}{\ifthenelse{\boolean{singlecolumn}}{\abscontentformatted}{\abscontent}}{}

\dropcap{T}he effects of a magnetic field on the spins of free electrons have been computed recently \cite{PLA2016,MAGMA2017} in the framework of quantum field theory (QFT), suggesting potential applications for spin-to-orbital angular momentum conversion in spintronic devices and electron vortex beams \cite{Ultramicro}. The work described here is motivated by the rich history in condensed matter and particle physics of extending quantum mechanical results by application of QFT. In particular, our study makes use of the fact that in QFT spin is precisely defined at the outset as a function of the quantum fields, rather than arising in quantum mechanics as an ad-hoc addition to the orbital angular momentum. Starting from the expression of the spin $\vec{S} = (S_1, S_2, S_3) = (S^{23}, S^{31}, S^{12})$ in terms of the fields,
\begin{equation} \label{Spin}
S^{ab} = \int d \mathbf{x} \, \psi^\dagger (\mathbf{x}) \, \frac{1}{2} \sigma^{ab} \, \psi(\mathbf{x}) = \int d \mathbf{x} \, \psi^\dagger (\mathbf{x}) \, \frac{i}{2} \gamma^a \gamma^b \, \psi(\mathbf{x}),
\end{equation}
the complete expression for the spin shift caused by the magnetic field \cite{PLA2016} is given by $\Delta_{\vec{A}} \vec{S} = |e| \int d\vec{x} \, \, [\vec{A} \times \vec{\rho}_E]$, where $\vec{A}$ is the magnetic potential, $\vec{\rho}_E = \frac{i}{m_e}\psi^\dagger \vec{\gamma} \psi$, and $\gamma^\mu=(\gamma^0,\vec{\gamma})$ defines the conventional Dirac matrices. We shall use the Weyl (chiral) basis in our presentation that follows.

\section*{Preliminary Details}
The purpose of the present letter is to study the effects of static electric and magnetic potentials on the energy of free electrons. With this aim, we shall begin with a free Dirac particle of mass $m_e$ described by a four-component wavefunction $\psi(\mathbf{x})$ satisfying the Dirac equation \cite{Dirac1},
\begin{equation}
(i\gamma^{\mu}\partial_{\mu}-m_e)\psi=0.
\end{equation}
The explicitly hermitian expression for the Hamiltonian of a free electron state, given in terms of the four components $\psi_1, \psi_2, \psi_3, \psi_4$ of the electron field, is thus
\begin{eqnarray}
\mathcal{H} = \int d\vec{x} && \left[ -\frac{i}{2}\overline{\psi} \gamma^{p} \partial_{p} + \frac{i}{2} (\partial_{p}\overline{\psi})\gamma^{p} + m_e \overline{\psi} \right] \psi \nonumber\\
= \int d\vec{x} && \left\{ 2m_e \, \Re\left[\, \psi_1^\ast \psi_3 + \psi_2^\ast \psi_4\,\right] \right. - \left.  \,\Im\left[\, \psi_1^\ast \partial_3 \psi_1 +\psi_1^\ast \left(\, \partial_1 -i\partial_2\,\right)\psi_2 \right.\right. +\left.\left. \psi_2^\ast \left(\, \partial_1 +i\partial_2\,\right)\psi_1 -\psi_2^\ast \partial_3 \psi_2  \,\right]\right.  \nonumber\\
&&+\left. \,\Im\left[\, \psi_3^\ast \partial_3 \psi_3 +\psi_3^\ast \left(\, \partial_1 -i\partial_2\,\right)\psi_4\right. \right. + \left.\left. \psi_4^\ast \left(\, \partial_1 +i\partial_2\,\right)\psi_3 -\psi_4^\ast \partial_3 \psi_4  \,\right]
\right.\}.
\label{ham}
\end{eqnarray}
It is clear above that the implied summation over $p$ indexes the three spatial components only. The introduction of an electromagnetic potential by the covariant four-vector $A_{\mu}=(A_0, \vec{A})$ will modify the free Hamiltonian following the minimal coupling prescription $\partial_{\mu} \rightarrow \partial_{\mu} - i |e| A_{\mu}$. From the modified Dirac equation, we derive
\begin{equation}
\Delta_{A_{\mu}}\psi=+\frac{|e|}{m_e}\gamma^\mu A_\mu \psi.
\end{equation}
Keeping only those terms to first order in the components of $A_{\mu}$, effectively replacing the four components $\psi_j \rightarrow \psi_j + \Delta_{A_{\mu}}\psi_j$ \cite{PLA2016} in the mass term of the integrand and using the minimal coupling replacement elsewhere in Equation (\ref{ham}), we find the energy shift due to the electromagnetic potentials:
\begin{equation}
\Delta_{A_{\mu}} \mathcal{H} = 2 |e| \int d\vec{x} \, \left\{A_0 \rho_0 + \vec{A} \cdot \left[\Re \left(-\psi_1^*\psi_2+\psi_3^*\psi_4\right), \Im \left(-\psi_1^*\psi_2+\psi_3^*\psi_4 \right), \frac{1}{2} \left(-|\psi_1|^2 + |\psi_2|^2 + |\psi_3|^2 - |\psi_4|^2 \right) \right] \right\},
\label{Deltaham}
\end{equation}
where $\rho_0 = \sum_{j=1}^4 |\psi_j|^2$. Further detail on this calculation can be found in the Methods.

By examination of the expressions in the dot product of Equation (\ref{Deltaham}), we arrive at a natural relationship between the electromagnetic energy shift and the spin. The spin current is defined generally as the quantity whose integral gives the spin vector itself, $\vec{S}=\int d\vec{x} \, \vec{s}$, and its components are given by
\begin{eqnarray}
\nonumber s_1=&& \Re\left(\,\psi_1^\ast \psi_2 + \psi_3^\ast \psi_4\, \right)\label{s1}\\ 
\nonumber s_2=&& \Im\left(\,\psi_1^\ast \psi_2 + \psi_3^\ast \psi_4\, \right)\label{s2}\\ 
s_3=&& \frac{1}{2}\left(\,|\psi_1|^2 - |\psi_2|^2 \right. + \left. |\psi_3|^2 - |\psi_4|^2 \,\right).
\label{s3}
\end{eqnarray}
If we transform these spin current components by sign reversal of the $\psi_1, \psi_2$ Dirac bilinear products, using the $\gamma^5 \equiv i\gamma^0\gamma^1\gamma^2\gamma^3$ and Pauli matrices $\sigma_p$, such that 
\begin{equation} \label{spintransform}
s_p^\prime=s^{\prime ab}=\frac{1}{2}\psi^\dagger\gamma^5\sigma^{ab}
\psi=\frac{1}{2}\psi^\dagger \begin{bmatrix}
-I_2 & 0 \\
0 & I_2
\end{bmatrix}
\begin{bmatrix}
\sigma_p & 0 \\
0 & \sigma_p
\end{bmatrix}
\psi=\frac{1}{2}\psi^\dagger \begin{bmatrix}
-\sigma_p & 0 \\
0 & \sigma_p
\end{bmatrix}
\psi,
\end{equation}
and define $s_0^\prime \equiv \psi^\dagger(\mathbf{x}) \, I_{4} \, \psi(\mathbf{x}) = \rho_0$, the expression for the energy shift becomes particularly compact:
\begin{equation}
\Delta_{A_\mu} \mathcal{H} = 2 |e| \int d\vec{x} \, A_\mu s^{\prime \mu}.
\label{sprime}
\end{equation}
Because we can use $\gamma^5$ to construct the chirality projection operators $\frac{1}{2}(I_4 \pm \gamma^5)$, Equation (\ref{spintransform}) constitutes a chiral transformation of the spin operator $\sigma^{ab}$.

A relevant feature of the obtained result is that such a magnetic energy shift is distinct from what one would expect in a quantum mechanical description. As is well known, this expression---called the Zeeman effect---can be written for an electron as a dot product between the electron angular momentum $\vec{J}$ and the magnetic field $\vec{B}$. For the contribution coming from the electron spin $\vec{S}$, we would have in quantum mechanics
\begin{equation}
\Delta _{\vec{A}} \mathcal{H}_{Zeeman} = -g_s \mu_B \vec{B} \cdot \vec{S},
\label{Zeeman}
\end{equation}
where $g_s \approx 2$ is the gyromagnetic factor and $\mu_B=|e|/2m_e$ is the Bohr magneton. One can see a resemblance between Equation (\ref{Zeeman}) above and the vector potential portion of the QFT scalar product in Equation (\ref{sprime}), with roughly a replacement of the magnetic field $\vec{B}$ with the magnetic potential $\vec{A}$ and likewise of the spin $\vec{S}$ with the transformed spin current $\vec{s}^{\,\,\prime}$. An analogous comparison could be made between the energy shift due to the electric potential $A_0$ and the Stark effect. 

We now want to compute the expectation value of the energy shift in Equation (\ref{sprime}) in a specific one electron state. We shall separate this energy shift into the two contributions $\Delta_{A_0} \mathcal{H}$ and $\Delta_{\vec{A}} \mathcal{H}$ from the electric and magnetic potentials, respectively. Consider the state with momentum $\vec{k}$ and defined as a linear combination of two spin eigenstates with complex coefficients:
\begin{equation} \label{Psi}
\left |\Psi(\vec{k}) \right \rangle = \lambda_+ \left|\uparrow, \vec{k} \right \rangle + \lambda_- \left|\downarrow,\vec{k}\right \rangle, 
\end{equation}
where the spin eigenstates are given by
\begin{equation}
\left|\uparrow, \vec{k} \right \rangle = \sqrt{2E} \, a_{\vec{k}}^{\uparrow\,\dagger} |0\rangle, \quad \left|\downarrow, \vec{k} \right \rangle = \sqrt{2E} \, a_{\vec{k}}^{\downarrow\,\dagger} |0\rangle, 
\label{spinup/down}
\end{equation}
with $E=\sqrt{m_e^2 + |\vec{k}|^2}$ and the state normalization fixed according to the prescription of Peskin and Schroeder \cite{Peskin} as
\begin{equation}
\braket{\uparrow, \vec{k}\,\,\,}{\uparrow, \vec{k}} = \braket{\downarrow, \vec{k}\,\,\,}{\downarrow, \vec{k}} = 2E (2\pi)^3 \delta^{(3)}(0).
\label{spinnorm}
\end{equation}
The normalization of our single electron state follows immediately:
\begin{equation}
\braket{\Psi(\vec{k})\,\,}{\Psi(\vec{k})} =  2E (2\pi)^3 \delta^{(3)}(0) \left(|\lambda_+|^2 + |\lambda_-|^2 \right).
\end{equation}

The starting expression for the expectation value of the energy shift is
\begin{equation}
\avg{\Delta_{A_\mu} \mathcal{H}}  = \frac{\bra{\Psi(\vec{k})}\Delta_{A_\mu} \mathcal{H} \ket{\Psi(\vec{k})}}{\braket{\Psi(\vec{k})}{\Psi(\vec{k})}}. \label{expavgH}
\end{equation}
To compute it, one needs to introduce the Fourier transforms of the electron fields, limiting the expressions to only those containing the fermionic creation and annihilation operators:
\begin{equation}
\begin{gathered}
\psi(\mathbf{x}) = \int \frac{d\vec{k}}{(2\pi)^3} \frac{1}{\sqrt{2E}} \sum_s \left(a^s_{\vec{k}}u^s e^{-i\mathbf{k} \cdot \mathbf{x}}   \right)\\ 
\overline{\psi}(\mathbf{x}) = \int \frac{d\vec{k}}{(2\pi)^3} \frac{1}{\sqrt{2E}} \sum_s \left(a^{s\dagger}_{\vec{k}} \overline{u}^s e^{i \mathbf{k} \cdot \mathbf{x}}\right),
\end{gathered}
\label{Fouriers}
\end{equation}
where $a^{s\dagger}_{\vec{k}}, a^s_{\vec{k}}$ are the creation and annihilation operators obeying the anticommutation relations $\left\{a^r_{\vec{k}}, a^{s\dagger}_{\vec{l}} \right\} = (2\pi)^3 \delta^{(3)}(\vec{k}-\vec{l} \,)\delta^{rs}$. The fermionic spinor fields $\overline{u}^s, u^s$ follow the choice of basis \cite{Peskin} in spin-$z$ eigenstates, such that
\begin{equation} \label{uzspinors}
u^{\uparrow}(\vec{k}) = (\sqrt{E-k_z}, 0, \sqrt{E+k_z}, 0), \quad u^{\downarrow}(\vec{k}) = (0, \sqrt{E+k_z}, 0, \sqrt{E-k_z}).
\end{equation}
To simplify the integration over the potentials $A_\mu$, we have assumed that its components can be approximated by their average values $\avg{A_0}, \avg{A_1}, \avg{A_2}, \avg{A_3}$ over the integration volume so that they can be extracted from the integral as numbers. This volume, in which these components are nonvanishing, is fixed by the scale $d$ of the experimental apparatus used to generate the electromagnetic fields. 

\section*{Results}
Each term in the expectation value (\ref{expavgH}) must be computed between four sets of bra-kets corresponding to the $\uparrow\uparrow, \downarrow\downarrow, \uparrow\downarrow$, and $\downarrow\uparrow$ configurations. Additional details on the organization and symmetry of these calculations are included in the Methods. The final expression for the magnetic energy shift in our single electron state is thus
\begin{equation}
\avg{\Delta_{\vec{A}} \mathcal{H}} = \frac{|\vec{k}|}{E} |e| A_3.
\label{finalavgH}
\end{equation}
This is a remarkable result: In the fully relativistic treatment, the first-order energy shift due to the magnetic potentials is completely independent of the spin-state coefficients $\lambda_\pm$. Furthermore, by choosing spinor fields corresponding to spin-$z$ eigenstates, only the $z$ component of the vector potential $\vec{A}$ survives in the expression for the average energy shift, due to symmetrical but cancelling contributions elsewhere (see Methods). In the ultra-relativistic limit ($E \approx|\vec{k}|$), this shift is proportional to the change ($|e|A_3$) in the conjugate momentum in the $z$ direction due to the introduction of magnetic potentials.

Choosing $\vec{A} = \frac{1}{2} \vec{B} \times \vec{x}$, and using characteristic values for a very weak bar magnet of $3 \times 10^{-4}$ tesla (3 gauss) and an experimental apparatus of dimension $d=1$ meter, we obtain $\left|\avg{\Delta_{\vec{A}} \mathcal{H}} \right| \lesssim 0.160$ MeV, just slightly more than $30\%$ of the electron's rest mass. This value is more than 20 times the maximum energy shift for slow electrons ($E \approx m_e$) with $(1-v^2)^{-1/2}=1.001$, for which $\left|\avg{\Delta_{\vec{A}} \mathcal{H}} \right|$ is about $1.4\%$ of the electron rest mass. As a comparison, the quantum mechanical Zeeman shift for these characteristic values is orders of magnitude smaller ($10^{-8}$ eV), as one might expect for such weak field strengths. We find these QFT estimates a good validation of the perturbative expansion employed, which only retains the magnetic field to first order. 

The expression for the average energy shift due to the electric potential is similar:
\begin{equation} \label{electricRL}
\avg{\Delta_{A_0} \mathcal{H}} = 2|e|A_0.
\end{equation}
This electric energy shift is similarly independent of the spin-state coefficients but also lacks information on the electron momentum, which can be understood at first order from the nature of the Lorentz force, $\partial_0 k_\mu = e (\partial_\mu A_\nu - \partial_\nu A_\mu) \partial_0 x^\nu$ or in three-vector notation $\vec{F}=e(\vec{E}+\vec{v}\times \vec{B})$. Computing the average energy shifts in the non-relativistic limit (NRL) where $\psi_1, \psi_2 \gg \psi_3, \psi_4$, we obtain
\begin{eqnarray} \label{electricNRL}
\avg{\Delta_{A_0} \mathcal{H}_{NRL}} =&& |e|A_0 \left(1- \frac{|\vec{k}|}{E} \frac{|\lambda_+|^2-|\lambda_-|^2}{|\lambda_+|^2+|\lambda_-|^2}\right), \\
\avg{\Delta_{\vec{A}} \mathcal{H}_{NRL}} =&& \frac{-|e|}{|\lambda_+|^2 + |\lambda_-|^2} \left\{\frac{m_e}{E} \left[A_1 \Re(\lambda_+^* \lambda_-) + A_2 \Im(\lambda_+^* \lambda_-) \right]+ \frac{1}{2} A_3 \left[|\lambda_+|^2 - |\lambda_-|^2 \right] \right\} + \frac{|e| A_3}{2}\frac{|\vec{k}|}{E}.
\label{magneticNRL}
\end{eqnarray}
Equations (\ref{electricNRL}, \ref{magneticNRL}) make apparent that, in the NRL, symmetries are broken which require the inclusion of spin-state information ($\lambda_\pm$) in the expressions for the average energy shift. Still, it is interesting to note that there exists a fixed term in both electric and magnetic shifts in the NRL that is entirely independent of the spin-state coefficients.

In the low-mass limit, there exists a correspondence equating the expression for chirality $\mathcal{X}$ to that of helicity:
\begin{equation} \label{chirality}
\mathcal{X} \xrightarrow{m_e \ll E} \frac{\vec{S} \cdotp \vec{k}}{\left| \,\vec{k} \,\right|}, \quad \mathcal{X}_{k_z} = S_z,
\end{equation}
where $\vec{S}$ is the spin defined by the Pauli matrices for a particle with momentum $\vec{k}=(0,0,k_z)$. Assuming a definition which reproduces the identification with helicity from Equation (\ref{chirality}), we see that the electric energy shift (\ref{electricNRL}) can be rewritten as
\begin{equation} \label{electrichiral}
\avg{\Delta_{A_0} \mathcal{H}_{NRL}} = |e|A_0 \left(1- \frac{2|\vec{k}|}{E} \avg{\mathcal{X}_{k_z}}\right).
\end{equation}
Thus, this average shift due to the electric potential in the NRL is a maximum for achiral states $\left(\avg{\mathcal{X}_{k_z}}=0\right)$, and attains a maximum value (equal to the fixed term $|e|A_0$) precisely half that of the fully relativistic result shown in Equation (\ref{electricRL}). Likewise, the fixed term $+\frac{|\vec{k}|}{E}|e|A_3/2$ in the average magnetic shift (\ref{magneticNRL}) is precisely half that of the relativistic shift (\ref{finalavgH}).

\begin{SCfigure}
\centering
\includegraphics[width=.5\linewidth]{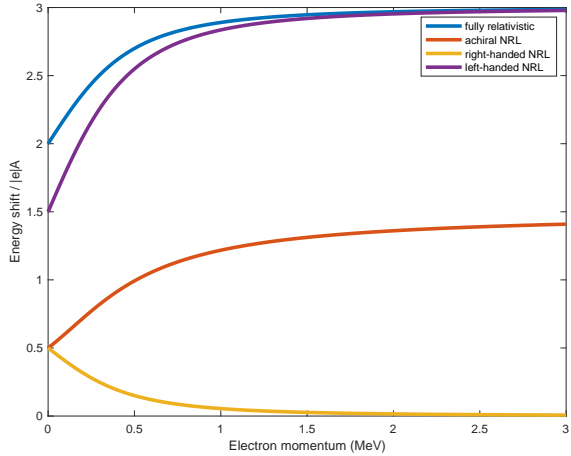}
\caption{\textbf{Expectation values of energy shifts due to electromagnetic potentials $A_\mu$.} The energy shift expectation values, computed in the single electron state $\left |\Psi(\vec{k}) \right \rangle = \lambda_+ \left|\uparrow, \vec{k} \right \rangle + \lambda_- \left|\downarrow,\vec{k}\right \rangle$, have been normalized by $|e|\bar{A}$, with $\bar{A} = A_0=A_1=A_2=A_3$. Dimensionless results for the fully relativistic treatment $\avg{\Delta_{A_\mu} \mathcal{H}}$ (blue), nonrelativistic limit (NRL) achiral state $\avg{\Delta_{A_\mu} \mathcal{H}_{NRL}^{\text{achir}}}$ (orange), and NRL completely polarized states $\avg{\Delta_{A_\mu} \mathcal{H}_{NRL}^{\text{pol, R}}}$ (yellow) and $\avg{\Delta_{A_\mu} \mathcal{H}_{NRL}^{\text{pol, L}}}$ (purple) are presented as functions of the electron momentum $|\vec{k}|$, in units of MeV.} 
\label{fig:polchirplot}
\end{SCfigure}

Putting our calculations for electric and magnetic potentials together, we obtain the fully general result 
\begin{equation}
\avg{\Delta_{A_\mu} \mathcal{H}} = |e|\left(2A_0+\frac{|\vec{k}|}{E}A_3\right)
\end{equation}
to first order, and for achiral electron states, we get
\begin{equation}
\avg{\Delta_{A_\mu} \mathcal{H}_{NRL}^{\text{achir}}} = |e| \left(A_0 -\frac{m_e}{2E}A_1 + \frac{|\vec{k}|}{2E}A_3 \right).
\end{equation}
For a completely polarized right- ($\lambda_-=0$) or left-handed ($\lambda_+=0$) electron state, $A_1$ and $A_2$ terms vanish:
\begin{equation}
\avg{\Delta_{A_\mu} \mathcal{H}_{NRL}^{\text{pol}}} = |e|\left[A_0 \left(1 \mp \frac{|\vec{k}|}{E} \right) + A_3 \left(\frac{|\vec{k}|}{2E} \mp \frac{1}{2} \right) \right].
\end{equation}
The difference between these energy shifts,
\begin{eqnarray} \label{polchirdiff}
\avg{\Delta_{A_\mu} \left(\mathcal{H}_{NRL}^{\text{pol}}- \mathcal{H}_{NRL}^{\text{achir}}\right)} =&& |e|\left(\mp \frac{|\vec{k}|}{E} A_0  + \frac{m_e}{2E}A_1 \mp \frac{1}{2} A_3 \right) \quad \text{and} \\
\label{poldiff}
\avg{\Delta_{A_\mu} \left(\mathcal{H}_{NRL}^{\text{pol,L}}- \mathcal{H}_{NRL}^{\text{pol,R}}\right)} =&& |e| \left(\frac{2|\vec{k}|}{E}A_0 + A_3 \right), 
\end{eqnarray}
can be experimentally measured to test the validity of our theory. From Figure (\ref{fig:polchirplot}), we can see that the latter difference in Equation (\ref{poldiff}) is larger than (\ref{polchirdiff}) for all nonzero values of $|\vec{k}|$. Indeed, $\avg{\Delta_{\bar{A}} \left(\mathcal{H}_{NRL}^{\text{pol,L}}- \mathcal{H}_{NRL}^{\text{pol,R}}\right)}$ is the sum of $\avg{\Delta_{\bar{A}} \left(\mathcal{H}_{NRL}^{\text{pol,L}}- \mathcal{H}_{NRL}^{\text{achir}}\right)}$ and $\avg{\Delta_{\bar{A}} \left(\mathcal{H}_{NRL}^{\text{achir}}- \mathcal{H}_{NRL}^{\text{pol,R}}\right)}$. This is consistent with what we would expect for achiral states, as they are intermediate between the extremes of completely right- and left-handed polarizations.

\section*{Discussion}
A key question that remains to be considered is the possible relevance of our results to macroscopic states in quantum optics, condensed matter, and biological physics, where collections of free or quasi-free electrons may be described in the formalism above. Indeed, though the description of such macroscopic states is complex, recent experimental studies \cite{Naaman} indicate that electrons transmitted through chiral molecules may be filtered according to their spin state. Furthermore, it has long been known that a sensitive dependence exists between the chirality of crystals and low-energy fluctuations introduced by perturbing the crystallization solution \cite{Sci1990}. 

Such sensitive relationships between biological function and chirality of the underlying spin state are apparent with both free and bound electron states. Several articles since 2005 have reported effects of weak magnetic fields on the rate of enzymatic synthesis of adenosine triphosphate \cite{HorePNAS} and oxidative species \cite{MartinoQB} by the flipping of electron spins in a quantum-coherent fashion. We have shown theoretically \cite{JTB} that palindromic DNA complexes of defined chirality are essential to the symmetric recruitment of energy by certain enzymes for the formation of synchronized DNA double-strand breaks. These evidences point to the existence of an elaborate hierarchy of order connecting the spin states of electron systems to their mesoscopic and macroscopic manifestations, across many orders of magnitude in the physical world. 

\matmethods{This work does not contain any experimental data or methods. All derivations and calculations were completed by hand. The figure was produced in MATLAB.

To derive Equation (\ref{Deltaham}), we proceed with the following replacements in the integrand of Equation (\ref{ham}), to first order in the potentials:
\begin{eqnarray}
&&2m_e \Re \left(\psi_1^* \Delta_{A_\mu}\psi_3 + (\Delta_{A_\mu}\psi_1^*)\psi_3 + \psi_2^* \Delta_{A_\mu}\psi_4 + (\Delta_{A_\mu}\psi_2^*)\psi_4 \right) \nonumber\\
&&-\Im \left[\psi_1^*(-i|e|A_3)\psi_1 + \psi_1^*(-i|e|A_1-|e|A_2)\psi_2+ \psi_2^*(-i|e|A_1+ |e|A_2)\psi_1 - \psi_2^*(-i|e|A_3)\psi_2 \right] \nonumber\\
&&+\Im \left[\psi_3^*(-i|e|A_3)\psi_3 + \psi_3^*(-i|e|A_1-|e|A_2)\psi_4+ \psi_4^*(-i|e|A_1+ |e|A_2)\psi_3 - \psi_4^*(-i|e|A_3)\psi_4 \right].
\end{eqnarray}

To organize the calculations for the expectation value of the energy shift, we consider the numerator of Equation (\ref{expavgH}), which requires evaluating four bra-kets for each term of the sandwiched operator expression. Starting with $\Delta_{A_0} \mathcal{H}$, we explicitly evaluate the bra-ket for the $\uparrow\uparrow$ configuration:
\begin{eqnarray} \label{explicitcalc}
&&2 |e| \int d\vec{x} \, A_0 |\lambda_+|^2 \bra{\uparrow, \vec{k}} \psi_1^* \psi_1 + \psi_2^*\psi_2 + \psi_3^*\psi_3 + \psi_4^*\psi_4 \ket{\uparrow, \vec{k}} \nonumber\\
&&= 2 |e| A_0 |\lambda_+|^2 \int d\vec{x} \int \frac{d\vec{p}\,d\vec{p}^{\,\prime}}{(2\pi)^6} \frac{e^{i(\vec{p}-\vec{p}^{\prime}) \cdot \vec{x}}}{\sqrt{4E_p E_{p^\prime}}} \left\langle \uparrow \left| \sum_{s,s^\prime} {a^{s^\prime}_{p^\prime}}^\dagger a^s_p \left[{u^{s^\prime}_1(p^\prime)}^*u^{s}_1(p)+{u^{s^\prime}_2(p^\prime)}^*u^{s}_2(p) +{u^{s^\prime}_3(p^\prime)}^*u^{s}_3(p) + {u^{s^\prime}_4(p^\prime)}^*u^{s}_4(p)\right]   \right| \uparrow \right\rangle \nonumber\\
&&= 2 |e| A_0 |\lambda_+|^2 \int \frac{d\vec{p}\,d\vec{p}^{\,\prime}}{(2\pi)^6} \frac{(2\pi)^3\delta^{(3)}(\vec{p}-\vec{p}^\prime)}{\sqrt{4E_p E_{p^\prime}}} \left\langle 0 \left| (2E_k) a^{\uparrow}_{\vec{k}} \sum_{s,s^\prime} {a^{s^\prime}_{p^\prime}}^\dagger a^s_p \left[\,\dotsb \,\right] {a^{\uparrow}_{\vec{k}}}^\dagger  \right| 0 \right\rangle \nonumber\\
&&= (2\pi)^3 2 |e| A_0 |\lambda_+|^2 \int d\vec{p} \, \frac{2E_k}{2E_p} \left\langle 0 \left| \delta^{(3)}(\vec{p}-\vec{k})\delta^{\uparrow s^\prime} \left[{u^{s^\prime}_1(p)}^*u^{s}_1(p)+{u^{s^\prime}_2(p)}^*u^{s}_2(p) +{u^{s^\prime}_3(p)}^*u^{s}_3(p) + {u^{s^\prime}_4(p)}^*u^{s}_4(p) \right] \delta^{(3)}(\vec{p}-\vec{k})\delta^{\uparrow s}  \right| 0 \right\rangle \nonumber\\
&&= (2\pi)^3 2 |e| A_0 |\lambda_+|^2 \delta^{(3)}(0) \left[{u^{\uparrow}_1(k)}^*u^{\uparrow}_1(k)+{u^{\uparrow}_2(k)}^*u^{\uparrow}_2(k) +{u^{\uparrow}_3(k)}^*u^{\uparrow}_3(k) + {u^{\uparrow}_4(k)}^*u^{\uparrow}_4(k) \right] \nonumber\\
&&= (2\pi)^3 2 |e| A_0 |\lambda_+|^2 \delta^{(3)}(0) \left[(E-k_z) + 0 + (E+k_z) + 0 \right] = \boxed{4 |e| A_0 E |\lambda_+|^2 (2\pi)^3 \delta^{(3)}(0)}, 
\end{eqnarray}
where in the last line we have employed the use of the spinor fields from Equation (\ref{uzspinors}). We see that the more general expression for these spinors along a fermion spin component axis with coordinates $\theta, \phi$ can be derived \cite{Peskin} from the two-component spinors
\begin{equation}
\xi(\uparrow)= \begin{pmatrix}
\cos \frac{\theta}{2} \\
e^{i\phi} \sin \frac{\theta}{2}
\end{pmatrix} , \quad \xi(\downarrow) = \begin{pmatrix}
-e^{-i\phi} \sin \frac{\theta}{2}\\
\cos \frac{\theta}{2}
\end{pmatrix}.
\end{equation}
By symmetry, we obtain a result similar to the boxed quantity (\ref{explicitcalc}) for the $\downarrow\downarrow$ configuration, with the replacement $\lambda_+ \rightarrow \lambda_-$. We get zero contributions from both opposite-spin configurations. Note that in $\Delta_{A_0} \mathcal{H}$ the normalization for our spin state $\Psi$ in the denominator of the expectation value precisely cancels the factor of $(|\lambda_+|^2 + |\lambda_-|^2)E (2\pi)^3\delta^{(3)}(0)$ contributed by the same-spin configurations.

Moving to $\Delta_{\vec{A}} \mathcal{H}$, we note that there are \textit{zero} contributions from the $A_1$ and $A_2$ terms, due to precise cancellation of contributions from the opposite-spin configurations, e.g.,
\begin{eqnarray}
&& \int d\vec{x} \, \bra{\uparrow, \vec{k}} \Re(-\psi_1^*\psi_2) \ket{\downarrow, \vec{k}} = - \int d\vec{x} \, \bra{\uparrow, \vec{k}} \Re(\psi_3^*\psi_4) \ket{\downarrow, \vec{k}} = -(2\pi)^3 \delta^{(3)}(0) \frac{m_e}{2}, \nonumber\\
&& \int d\vec{x} \, \bra{\uparrow, \vec{k}} \Im(-\psi_1^*\psi_2) \ket{\downarrow, \vec{k}} = - \int d\vec{x} \, \bra{\uparrow, \vec{k}} \Im(\psi_3^*\psi_4) \ket{\downarrow, \vec{k}} = +(2\pi)^3 \delta^{(3)}(0) \frac{im_e}{2},
\end{eqnarray}
and nothing from the same-spin configurations. By symmetry with the $A_0$ bra-kets computed above, we can easily find the $A_3$ terms as expressed in the following relations:
\begin{eqnarray}
&& \int d\vec{x} \, \bra{\uparrow, \vec{k}} -|\psi_1|^2 + |\psi_2|^2 + |\psi_3|^2 -|\psi_4|^2 \ket{\uparrow, \vec{k}} = +(2\pi)^3 \delta^{(3)}(0) 2 |\vec{k}| = \int d\vec{x} \, \bra{\downarrow, \vec{k}} -|\psi_1|^2 + |\psi_2|^2 + |\psi_3|^2 -|\psi_4|^2 \ket{\downarrow, \vec{k}}. 
\end{eqnarray}
Therefore the total contributions to $\Delta_{\vec{A}} \mathcal{H}$ in our spin state $\Psi$ \textit{all} come from the $A_3$ terms, with a similar cancellation of a factor of $(|\lambda_+|^2 + |\lambda_-|^2) (2\pi)^3\delta^{(3)}(0)$ by the fixed normalization in the denominator of the expectation value.
}

\showmatmethods{} 

\acknow{PK would like to acknowledge discussions with C. Verzegnassi (University of Udine and Association for Medicine and Complexity, Italy) and support from the US-Italy Fulbright Commission and the Whole Genome Science Foundation.}

\showacknow{} 

\bibliography{pnas-sample}

\end{document}